\documentclass[aps,pre, amsmath,amssymb,twocolumn,groupedaddress]{revtex4-1}

\usepackage{bm}
\usepackage{graphicx}
\usepackage{amsfonts}
\usepackage{amssymb}
\usepackage{wasysym}
\usepackage{amsthm}
\usepackage{MnSymbol}%
\usepackage{color}

\begin{document}

\title{Dynamical Clustering Interrupts Motility Induced Phase Separation in Chiral Active Brownian Particles}

\author{Zhan Ma}
\author{Ran Ni}
\email{r.ni@ntu.edu.sg}
\affiliation{%
 School of Chemical and Biomedical Engineering, Nanyang Technological University, 637459, Singapore}

\begin{abstract}
One of the most intriguing phenomena in active matter has been the gas-liquid like motility induced phase separation (MIPS) observed in repulsive active particles. However, experimentally no particle can be a perfect sphere, and the asymmetric shape, mass distribution or catalysis coating can induce an active torque on the particle, which makes it a chiral active  particle. Here using computer simulations and dynamic mean-field theory, we demonstrate that the large enough torque of circle active Brownian particles (cABPs) in two dimensions generates a dynamical clustering state interrupting the conventional MIPS. Multiple clusters arise from the combination of the conventional MIPS cohesion, and the circulating current caused disintegration. The non-vanishing current in non-equilibrium steady states microscopically originates from the motility ``relieved'' by automatic rotation, which breaks the detailed balance at the continuum level. This suggests that no equilibrium-like phase separation theory can be constructed for chiral active colloids even with tiny active torque, in which no visible collective motion exists. This mechanism also sheds light on the understanding of dynamic clusters observed in a variety of active matter systems.
\end{abstract}

\maketitle
\section*{Introduction}
Active matter can spontaneously form structures not restricted by equilibrium thermodynamics, as they keep dissipating energy and break the time-reversal symmetry locally.
{For example, while the gas-liquid like phase separation in equilibrium requires cohesive interactions~\cite{vdwthesis}, active colloids can undergo motility-induced phase separation (MIPS) resulting from the combination of self-propulsion and steric repulsion.}
Essential physics of MIPS has been well captured by linear swimmer models like active Brownian particles (ABPs) \cite{PhysRevLett.108.235702, PhysRevLett.110.055701}, run-and-tumble particles (RTPs) \cite{PhysRevLett.100.218103, Cates_2013} and active-Ornstein-Uhlenbeck particles (AOUPs) \cite{PhysRevE.103.032607}.
The coarse graining of those microscopic linear models usually restores the detailed balance at the continuum level and MIPS was mostly understood based on equilibrium-like phase separations~\cite{cates2015motility,ma2020,Solon_2018}.

Despite the success of linear swimmer models in capturing the essential physics of MIPS, they fail in describing the commonly observed chiral swimmers.
Many biological microorganisms swim in circular and helical trajectories near surface or in chemical gradients \cite{bohmer2005ca2+, taktikos2011modeling}, including bacteria \cite{diluzio2005escherichia, di2011swimming}, sperm cells, and some alga \cite{martinez2012differential}.
Those chiral swimming patterns play an important role in surface selection, attachment and forming microcolonies of microorganism \cite{utada2014vibrio}.
It is also convenient to design synthetic circle swimmers by asymmetric shape \cite{gibbs2009design, gibbs2011geometrically, kummel2013circular}, mass distribution \cite{campbell2017helical}, catalysis coating \cite{archer2015glancing}, to control the radius and frequency of circular trajectories.
Individually, these chiral swimmers exhibit intriguing phenomena like gravitaxis \cite{ten2014gravitaxis}, abnormal transportation \cite{ManoE2580, C8SM02030B, RevModPhys2016}, etc.

\begin{figure*}[!t]
\centering
\includegraphics[width=1.0\linewidth]{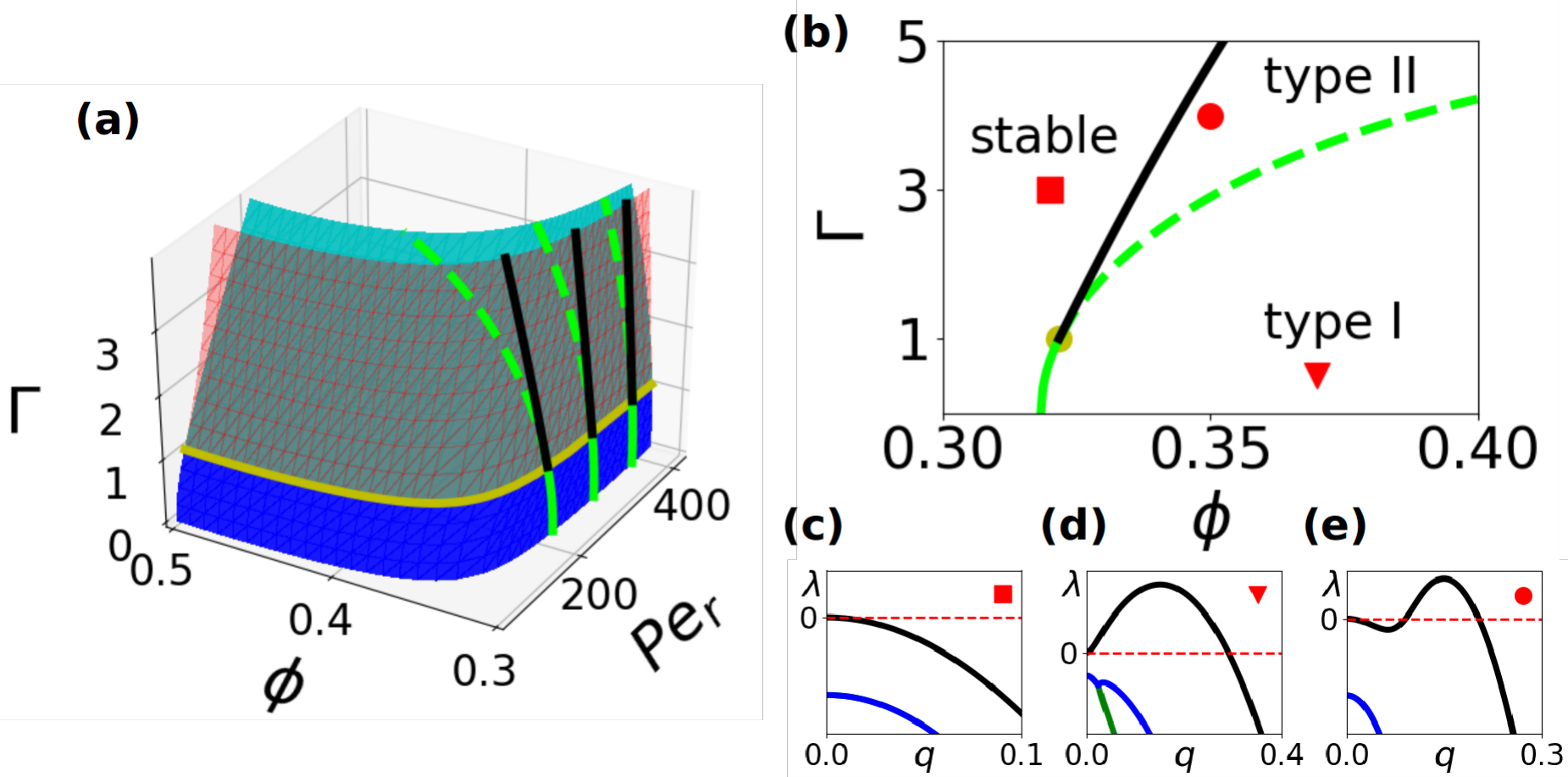}
\caption{\label{fig:instability} \textbf{Theoretically predicted phase diagram of circle active Brownian particles.} (a) Predicted phase boundaries for control parameter $[\phi,\mathrm{Pe}_r,\Gamma]$ at $D_r=0.1$, with phenomenological mean-field parameter $D=151 \sigma^2 D_r$, $\phi^*=0.63$. Blue plane separates the stable homogeneous state with type I instability. Red plane separates the stable homogeneous state with type II instability, and the cyan plane is pseudo separation plane for type I and II instability. {The green/black curves are the cross section of type I/II phase plane with fixed $\mathrm{Pe}_r$ plane. The yellow curve marks $\Gamma=1$.}  (b) Cross section at $\mathrm{Pe}_r=240$. (c{-e}) The dispersion 
relationship $\lambda(q)$ for stable homogeneous state ($\blacksquare$), type I instability ($\blacktriangledown$), type II instability ($\bullet$).}
\end{figure*}

The emergent behaviour of circle swimmers with explicit alignments was studied by generalizing the Vicsek model, where slow rotations enhance the polarization in macroflocks, while fast rotations induce secondary instabilities leading to phase synchronized microflocks \cite{liebchen2017collective}.
However, these flocks are integrated by phase-lock mechanism and suffer from strong fluctuations once including excluded volume interactions, which occur naturally among circle swimmers \cite{levis2018micro}.
In simulations of purely repulsive circle swimmers, significant suppression of MIPS, antiwise rotation of macrodroplet at some optimal slow rotation frequency \cite{liao2018clustering}, {formation of vortex arrays structure \cite{PhysRevE.87.032712}}, and collective oscillation of density under strong chirality \cite{liu2019collective}
were observed.
Recently, a field theory quantitatively accounting for the suppression of MIPS was proposed~\cite{bickmann2020analytical}.
However, as they consider the effect of self-propelling torque using an effective rotational diffusion coefficient, it is only valid in slow rotation region.
In contrast, non-equilibrium hyperuniform fluids were found in a deterministic chiral active particle model \cite{lei2019nonequilibrium,lei2019nonequilibrium2}, in which the active rotation rate can be seen as extremely fast, as the rotational noise is zero.
Therefore, the unified understanding of the collective assembly of chiral active swimmers remains unknown.
To this end, we formulate a hydrodynamic description from the microscopic dynamics of circle active Brownian particles (cABPs), and it is valid both at slow and fast rotations.
We find a short wavelength instability leading to a dynamical clustering state at fast rotation, and the theoretically predicted phase boundary well agrees with large-scale Brownian dynamics simulations.
By discussing the hydrodynamic matrix, we propose a novel instability mechanism originating from the combination of conventional MIPS cohesion and the circulating current caused disintegration.
We also verify that the system spanning circulating current originates from the motility ``relieved'' by fast rotation from temporal fluctuations.
Our results may help understand the general mechanism of self-limiting size cluster formation without introducing explicit alignment to synchronize phases \cite{liebchen2017collective} or chemical signalling \cite{PhysRevLett.108.268303, PhysRevLett.115.258301}.
More generally, non-vanishing current in steady states is a distinct feature of non-equilibrium systems, as not restricted by the thermodynamic balance.
Our results suggest that no equilibrium-like theory, based on detailed balance, can be constructed to understand chiral active colloids.

\section*{Model of Circle active Brownian particles}
We consider $N$ self-propelled particles in two-dimensions interacting via the repulsive, pairwise additive, Weeks-Chandler-Andersen potential:
\begin{equation}
\label{EqWCA}
	V(r)=4\epsilon \left[ \left(\frac{\sigma}{r} \right)^{12} - \left(\frac{\sigma}{r} \right)^6 \right]+\epsilon,
\end{equation}
with the cut-off at $r_c=2^{1/6}\sigma$, beyond which $V=0$. 
Here $\sigma$ is the nominal particle diameter, $\epsilon$ determines the interaction strength, and $r$ is the center-to-center distance between two particles. 
Particle $i$ with positions $\mathbf{r}_i$ and orientations $\mathbf{e}_i=(\cos \theta_i, \sin \theta_i)$ evolves in response to a systematic driving force and rotational torque, according to the overdamped Langevin equations~\cite{ma2017driving}:
\begin{equation}
\begin{aligned}
	\dot{\mathbf{r}}_i &=-\mu \sum _{i\neq j} \nabla _i V(|\mathbf{r}_i-\mathbf{r}_j|)+v_0 \mathbf{e}_i+\sqrt{2D_t} \bm{\xi}_i,\\
	\dot{\theta}_i &=\omega_0 +\sqrt{2D_r}\nu_i.
\end{aligned}
\end{equation}
Here, $\mu$ is the translational mobility and $v_0$ is the magnitude of the self-propulsion velocity.
The Gaussian white noise $\bm{\xi}_i$ models the interaction with the solvent, but as the translational diffusion effect is much smaller than that of the self-propulsion, we set $D_t=0$. 
To model the circular motion of active particles, the orientation $\theta_i$ displays a drift of constant angular velocity $\omega_0$ alongside the rotational diffusion coefficient $D_r$, with $\nu_i$ the unit Gaussian white noise $\langle \nu_i(t)\nu_j(t')\rangle=\delta_{ij}\delta(t-t')$. 
In the dilute and deterministic limit, individual particle moves in a circular trajectory with revolution radius $R=v_0/\omega_0$.
Then, there are three characteristic time scales: rotational diffusion $\tau_r=1/D_r$, circular motion $\tau_{\omega}=2\pi/\omega_0$ and ballistic motion $\tau_b=1/(v_0\sqrt{\rho_0})$. 
To construct the full phenomenology of this model requires scanning a four-parameter phase diagram, with space and time units $\sigma$ and $1/D_r$, respectively, parametrized by the P$\acute{\text{e}}$clet number $\mathrm{Pe}_r=v_0/(\sigma D_r)$, the dimensionless angular velocity $\Gamma=\omega_0/D_r$, the potential stiffness $\kappa=\mu \epsilon/ (v_0 \sigma)$, and the area fraction $\phi=\pi \sigma^2N/(4L^2)=\rho_0 \pi \sigma^2/4$ with $L$ and $\rho_0$ the side length of the simulation box and density, respectively. 
Here, we set $\sigma=1$, $\mu=1$, and fix $1/\kappa=v_0/\epsilon \equiv 24$, so that two head-to-head colliding cABPs reach the dynamic balance between their self-propulsion and pairwise interaction.
{In our Brownian dynamics simulations, periodic boundary conditions are applied in both directions, and the time step $\Delta t$ satisfies $v_0 \Delta t < 10^{-3}\sigma$. To make ensure the system reaching a steady state, we calculate the total potential energy of the system, which is ensured to be stable for at least $1000\tau_b$ before the measurement of any quatity in the steady state.}

\section*{Results}
\subsection*{Hydrodynamic description and finite wavelength instability}
To understand the emergence of inhomogeneous states, with the mean-field approximation\cite{bialke2013microscopic,speck2014effective,speck2015dynamical}, we obtain the dynamic equations for the density and polarization fields $\rho(\mathbf{r})=\sum_{i} \delta (\mathbf{r}-\mathbf{r}_i)$, $\mathbf{W}(\mathbf{r})=\sum_i \delta (\mathbf{r}-\mathbf{r}_i) \mathbf{e}_i$, respectively, as 
\begin{equation}
\label{eq:Hydro}
\begin{aligned}
	\dot{\rho}=& -\nabla \cdot (v_\text{e} \mathbf{W})+D\nabla^2 \rho,	\\
	\dot{\mathbf{W}}=& -\frac{1}{2}\nabla(\rho v_\text{e})+(D\nabla^2-D_r)\mathbf{W}+{\omega_0\hat{\mathbf{z}}}\times \mathbf{W}.
\end{aligned}
\end{equation}
Here, $v_\text{e}(\rho)=v_0-\zeta \rho$ is the mean-field effective velocity with $\zeta$ reflecting how strongly the motility is slowed down by its neighbours, and it can be written as $v_\text{e}(\phi)=v_0(1-\phi/\phi^*)$,  with $\phi^*$ the packing fraction damping $v_e$ to zero. 
{$\hat{\mathbf{z}}$ is the unit vector indicating the direction of torque.}
Although this linear relationship was originally proposed for linear ABPs, we verify that it is still valid in homogeneous cABPs, and $\phi^*$ does not depend on $\omega_0$ (see SM).
$D$ describes the effective diffusion caused by particle collisions.

\begin{figure*}[!t]
\centering
\includegraphics[width=1.0\linewidth]{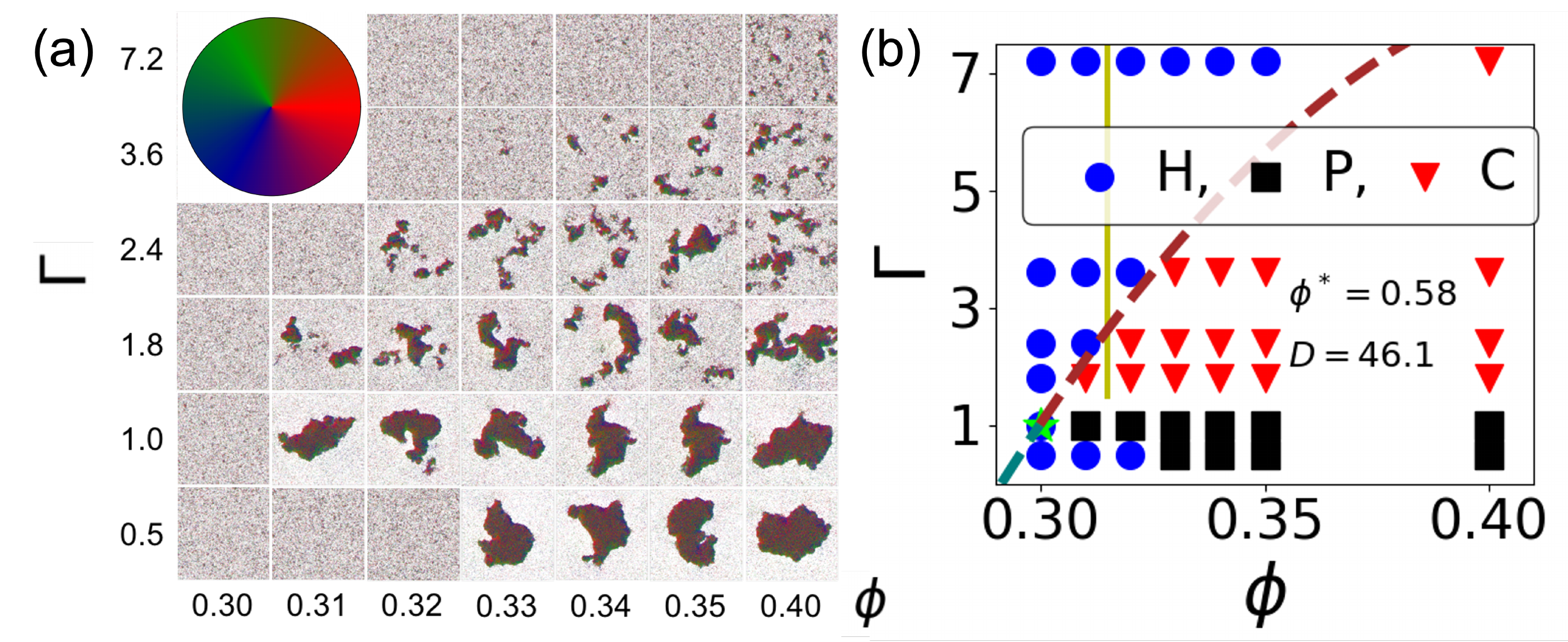}
\caption{\label{fig:DCstate} \textbf{Simulated phase diagram of circle active Brownian particles.} (a) Typical snapshots of three states, i.e. homogenous state (H), phase separated state (P), and dynamical clustering state (C). Color of particles indicates the {self-propulsion} orientation {shown in the top-left inset}. (b) Simulated hase diagram of cABPs, in which blue dots, black squares, and red triangles represent the homogeneous state (H), the MIPS state (P), and the dynamic clustering state (C), respectively. The red/cyan dashed curves are the theoretically predicted phase boundary between the stable homogeneous state and type II/I instability, with fitting parameter $\phi^*=0.58$, $D=461\sigma^2 D_r$. Here $D_r=0.1, \mathrm{Pe}_r=360$. The yellow line is {the} H-C phase boundary obtained in simulations for $\mathrm{Pe}_r=1200$ and $D_r=0.1$.}
\end{figure*}

Apparently, the isotropic homogeneous steady state $H(\rho,\mathbf{W})=(\rho_0,\bm{0})$ is a solution for \eqref{eq:Hydro}.
The linearized dynamics of fluctuations are
\begin{equation}
\label{eq:hydromatrix}
		\begin{bmatrix}
			\lambda+\Gamma_{\rho} & i \bar{v}_e q & 0 \\
			i\eta q & \lambda+\Gamma_r & \omega_0	\\
			0 & -\omega_0 & \lambda+\Gamma_r
		\end{bmatrix}
		\begin{bmatrix}
			\tilde{\delta \rho}\\ \tilde{w}_{\|}\\ \tilde{w}_{\perp}
		\end{bmatrix}=0,
\end{equation}
where {the Fourier transform is defined as $\tilde{f}=\int dt d\mathbf{r} f\exp(-\lambda t-i\mathbf{q}\cdot \mathbf{r})$. } $\tilde{w}_{\|/\perp}$ is the polarization fluctuation longitudinal/transversal to $\mathbf{q}$. Damping rates are $\Gamma_{\rho}=Dq^2$, $\Gamma_r=Dq^2+D_r$.
Density and longitudinal polarization evolutions are coupled by the average effective velocity $\bar{v}_{e}=v_0-\zeta \rho_0$ and effective compressional modulus $\eta=(\bar{v}_e-\zeta \rho_0)/2$.
Intuitively, the positive $\eta$ drives the current to avoid the denser zone, while the negative $\eta$ results in the current pointing to the denser zone.
MIPS instability in linear ABP models is usually intrigued by $\partial v_e/\partial \rho<0$ and $\eta<0$ suggesting that particles mobility slows down by the steric repulsion from its neighbours, and particles tend to accumulate where they move slower \cite{Solon_2018}.
Such slowdown-accumulation feedback loop provides an effective cohesion for phase separation, and we verify here that this mechanism is also preserved in chiral swimmers.
However, the transversal component $\tilde{w}_{\perp}$ cannot be decoupled from the other two modes at the linear order as in normal fluids due to the convection induced by automatic rotation $\omega_0$.

Looking at the wave mode $\exp(\lambda t+i\mathbf{q}\cdot \mathbf{r})$, the dispersion relation $\lambda(q=|\mathbf{q}|)$ unveils two types of instability of homogeneous state depending on the relative strength of angular velocity and rotational diffusion $\Gamma=\omega_0/D_r$.
For slow rotation $\Gamma<1$, increasing $\phi$ and $\mathrm{Pe}_r$ leads to the MIPS like instability (Fig.~\ref{fig:instability}(d)) when $\phi>\phi^{\text{pb}}_\text{I}=(3-\sqrt{1-16D/(v_0^2\tau_{\text{er}})})\phi^*/4$, where $\tau_{\text{er}}=D_r/(D_r^2+\omega_0^2)$ accounts for the suppression of MIPS by automatic rotation similar to Ref~\cite{bickmann2020analytical}.
Whereas for fast rotation $\Gamma>1$, we find a qualitatively different picture, where an instability starts at $q>q^*>0$ (Fig. \ref{fig:instability} (e)) corresponding to the short wavelength fluctuation.
This type II instability arises when $\phi>\phi^{\text{pb}}_\text{II}=(3-\sqrt{1-32D\omega_0/v_0^2})\phi^*/4$. 
As $\phi^{\text{pb}}_\text{I}>\phi^{\text{pb}}_\text{II}$ in this regime, the MIPS instability is interrupted (with pseudo-phase boundary shown by the cyan plane in Fig.~\ref{fig:instability} (a) and dashed green curves in Fig.~\ref{fig:instability} (b)).

\subsection*{Dynamical clustering state}
\begin{figure}[!b]
\centering
\includegraphics[width=1.0\linewidth]{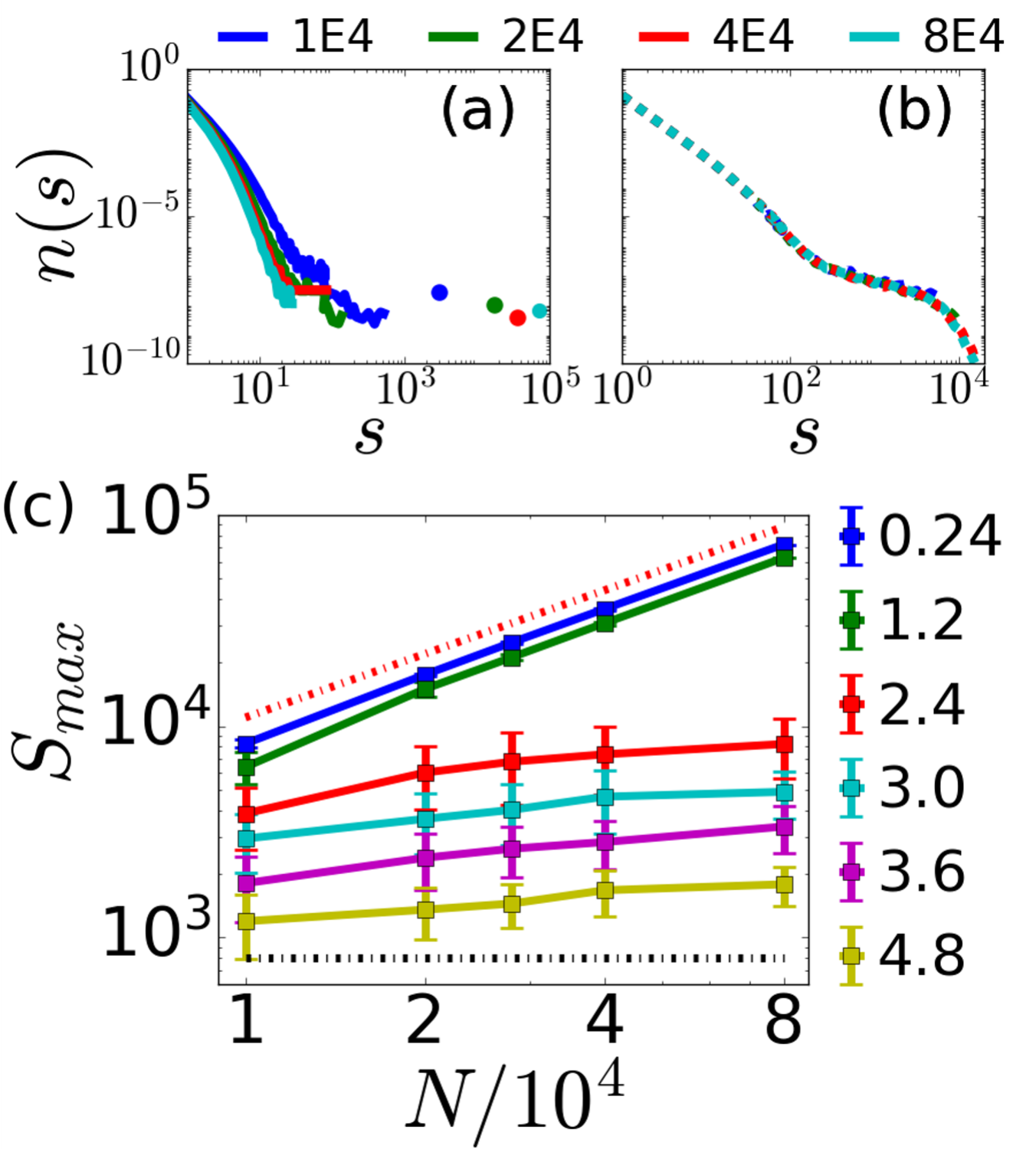}
\caption{\label{fig:FSscale} \textbf{Finite size analysis for cluster size distribution.} (a) $n(s)$ at $\Gamma=0.24$ for different number of particles $N$ in the system indicated by colors. (b) $n(s)$ at $\Gamma=2.4$. (c) Maximum cluster size $S_{\text{max}}$ {as} a function of system size $N$, colors indicate $\Gamma$. The red dashed line indicates $\propto N$, and the black dashed line represents a constant. Here $D_r=0.1, \mathrm{Pe}_r=240, \phi=0.40$.}
\end{figure}

To specify different inhomogeneous states from two types of instability, we simulate the collective behaviour of $N=40,000$ identical cABPs with fixed $\mathrm{Pe}_r$ and increasing $\phi$, and the typical snapshots are summarized in Fig.~\ref{fig:DCstate}(a).
For slow rotations $\Gamma \lesssim1$, we reproduce the MIPS as in ABPs: the system remains homogeneous (state H) below the spinodal ($\phi<\phi^{\text{pb}}_{\text{I}}$), whereas $\phi > \phi^{\text{pb}}_{\text{I}}$ induces a bulk phase separation with a single dense liquid droplet coexisting with a dilute gas phase (state P).
At $\phi > \phi^{\text{pb}}_{\text{II}}$ with fast rotation $\Gamma \gtrsim 1$, we observe the emergence of multiple dynamic clusters (state C), and they continuously merge, split and decay, but never merge into a stable dense bulk phase~\cite{PhysRevLett.125.178004,PhysRevLett.125.168001} (SM Video).
{Though there are small temporary clusters emitted out from the bulk dense phase in MIPS, they are negligible compared with the system size.
The finite size scaling is studied in Fig.~\ref{fig:FSscale} to characterize the states.}
The H-C phase boundary from computer simulation agrees with the linear instability predicted $\phi^{\text{pb}}_{\text{II}}$ with fitting parameter $\phi^*=0.58$ and $D=461\sigma^2 D_r$ as shown by the red dashed curve in Fig.~\ref{fig:DCstate}(b).
The green dot marks $\Gamma=1$, below which the predicted $\phi^{\text{pb}}_{\text{I}}$ shown by cyan line deviates from the H-P phase boundary in simulation, and this may result from the large hysteresis loop in MIPS.

To further examine the phase boundary predicted in the hydrodynamic theory, we derive the slope of H-C phase boundary in the parameter space $[\phi,\Gamma]$
\begin{equation}
\label{eq:pbslope}
	\frac{\partial \Gamma}{\partial \phi^{\text{pb}}_{\text{II}}}=\frac{v_0^2}{4DD_r\phi^*}\left( 3-4\frac{\phi}{\phi^*} \right),
\end{equation}
which suggests that the slope of type II instability increases with $v_0^2$. 
Thus we increase $v_0$ by $3$ times in simulation and do observe the gentle increasing boundary (the red dashed curve) changing to nearly vertical as shown by the yellow line in Fig.~\ref{fig:DCstate}(b).
The agreement between simulations and the instability analysis verifies that the dynamical clustering state is induced by the short wavelength instability at fast rotation. 

\begin{figure}[!t]
\centering
\includegraphics[width=1.0\linewidth]{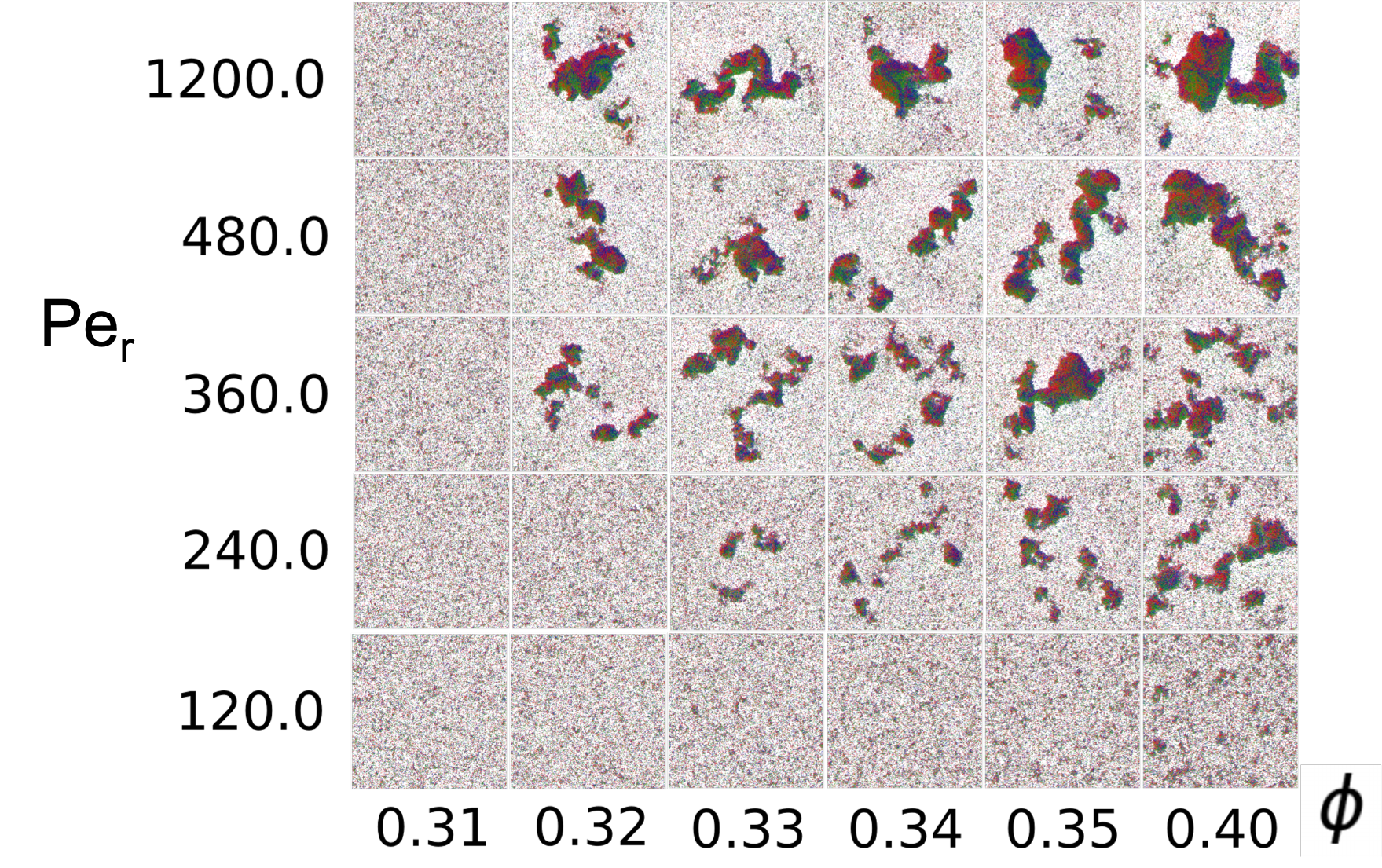}
\caption{\label{fig:pDCstate} \textbf{Typical snapshots of dynamical clustering state.} $\Gamma=2.4$.}
\end{figure}

To eliminate the possibility that the dynamical clustering can be a metastable state or near-critical fluctuation of the bulk phase separation, we further simulate the system deep in the dynamical clustering state of $[\phi,\text{Pe}_r]$ for $\Gamma>1$ as shown in Fig.~\ref{fig:pDCstate}.
Although clusters start emerging at $\phi \simeq 0.32$, $D_r = 0.1$ and $\mathrm{Pe}_r \simeq 240$, with further increasing $\phi$ to $0.40$ and $\mathrm{Pe}_r$ by a factor of $5$, these clusters  remain highly dynamical instead of merging into a stable giant cluster as in MIPS. {To ensure that the systems have reached the dynamical clustering steady states, we also perform simulations starting from MIPS configurations, in which the single giant droplet splits into multiple smaller clusters after switching on fast rotation.}
This suggests that the dynamical clustering is a well-defined non-equilibrium steady state.

As active matter systems generally suffer from strong finite-size effects \cite{PhysRevLett.92.025702}, we further examine the effect of system size in two inhomogeneous states as shown in Fig.~\ref{fig:FSscale}.
$n(s)$ is the cluster size distribution (CSD) for cluster containing $s$ particles.
{Particles with distance less than $r_c$ are considered to be in the same cluster.}
The homogeneous state exhibits a CSD decaying in the form $n(s)\propto s^{-\alpha} \exp(-s/s_{\xi})$ with $\alpha \simeq 1.9$ (see SM).
In MIPS, a separated dot represents the single giant cluster (Fig.~\ref{fig:FSscale}(a)), and the time averaged largest cluster size $S_{\text{max}}$ increases linearly with system size confirming the bulk phase separation scenario.
For the dynamical clustering state, a pronounced bump emerges abruptly at an intermediate size $s\sim 10^3-10^4$ after crossing $\phi^{\text{pb}}_{\text{II}}$.
Remarkably, increasing the system size by $8$ times does not produce any visible change in the CSD of dynamical clustering state.
Besides, $S_{\text{max}}$ increases much slower than $\propto N$ for dynamical clustering state.
The gradual slope may result from the emergence of temporal extremely large cluster in fluctuation as the system size increases.
These confirm that the characteristic length scale is independent of the system size $N$, while determined by its self-propulsion parameters, consistent with the predicted short wavelength instability.

\subsection*{Circulating current caused disintegration}
As the conventional MIPS cohesion effect is preserved in the linear order hydrodynamics of chiral swimmers, to form dynamic clusters, there should be a disintegration mechanism.
Thus we explore the steady state current, which vanishes in equilibrium systems as a result of the balance of chemical potential and pressure everywhere {\cite{PhysRevX.8.031080}}.
For slowly varying fields, we neglect the temporal derivative as well as the viscosity term in \eqref{eq:Hydro} to obtain the adiabatic solution of the polarization field
\begin{equation}
\label{eq:adiabaticPolar}
	\mathbf{W}_{\text{ad}}=-\frac{1}{2(D_r^2+\omega_0^2)} 
		\begin{bmatrix}
			D_r & -\omega_0	\\
			\omega_0 & D_r
		\end{bmatrix}
		\nabla j(\rho),
\end{equation}
where $j(\rho)=\rho v_e(\rho)$ represents the scalar strength of effective current.
As $\mathbf{J}_{\text{ad}}=v_e \mathbf{W}_{\text{ad}}$, we find a non curl free current in steady states of chiral swimmers, {$\nabla \times \mathbf{J}_{\mathrm{ad}}=\omega_0\left[\zeta(\eta+v_e)|\nabla \rho|^2-\eta v_e\nabla^2\rho \right]/(D_r^2+\omega_0^2)$.
The circulating current originates from the mixture of spatial gradients induced by automatic rotation. 
Intuitively, such system spanning circulating current can continuously disassemble the dynamic clusters.
Different from the equilibrium-like MIPS in ABPs, the steady state current in cABPs is a distinct feature of broken detailed balance.

\begin{figure}[!t]
\centering
\includegraphics[width=1.0\linewidth]{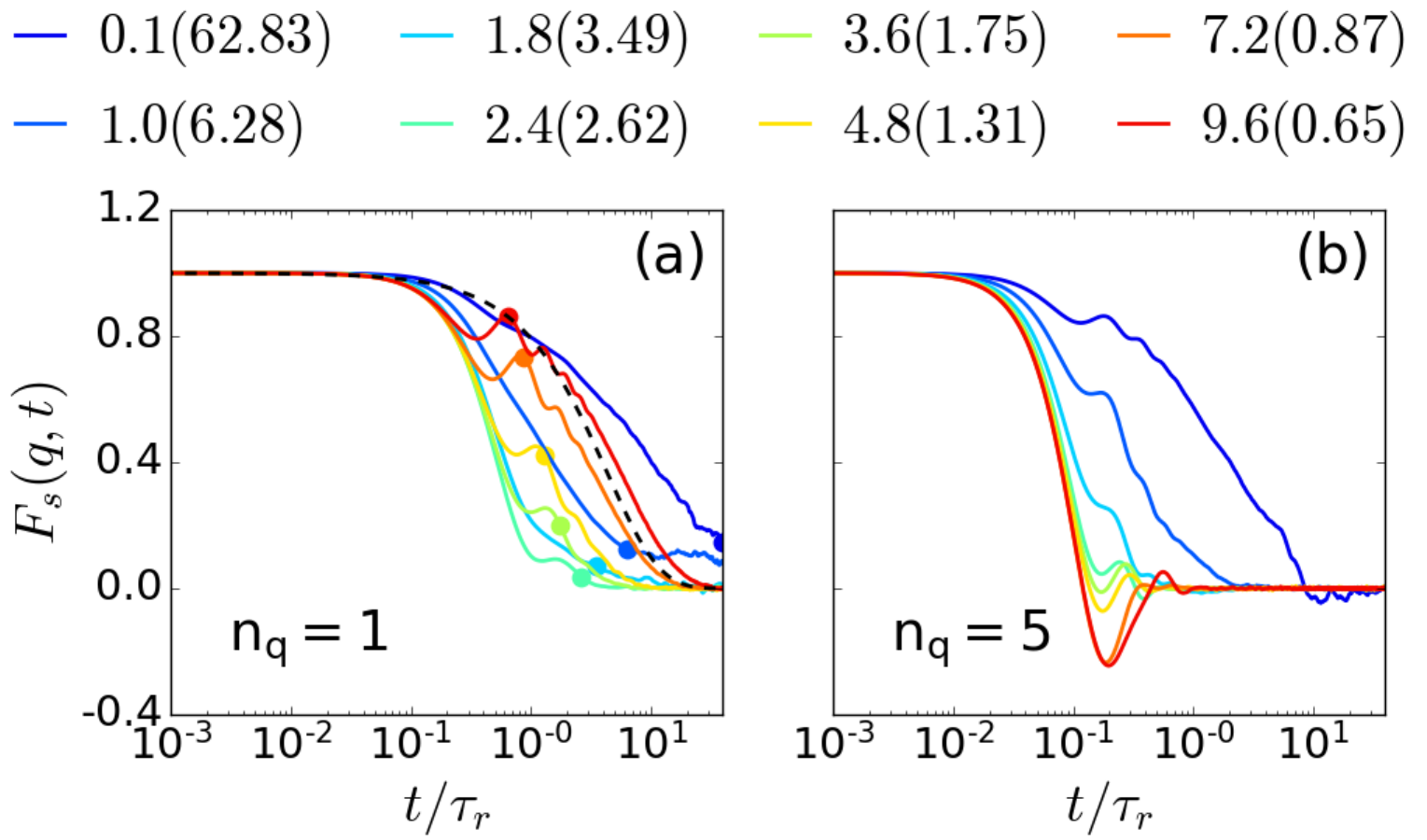}
\caption{\label{fig:sISF} \textbf{Self intermediate scattering function of circle active Brownian particles.}  (a) $n_q=1$, the dots indicate the time scale $\tau_{\omega}/\tau_r$. (b) $n_q=5$, the color indicates $\Gamma$ and the inside bracket is $\tau_{\omega}/\tau_r$. Here $D_r=0.1$, $\phi=0.40$, $\mathrm{Pe}_r=360$.}
\end{figure}

To understand the microscopic origin of circulating current, we measure the self intermediate scattering function (sISF) which reveals dynamic patterns at different spatiotemporal regimes,
\begin{equation}
\label{Eq:sISF2}
	F_s(q=|\mathbf{q}|,t)=\frac{1}{N}\displaystyle{\sum_{i=1}^N}\langle \exp(i\mathbf{q}\cdot [\mathbf{r}_i(t)-\mathbf{r}_i(0)]) \rangle.
\end{equation}
Here $\langle \cdot \rangle$ represents the ensemble average, and we define $n_q = qL/(2\pi)$.
{The sISF of individual circle swimmer was studied in \cite{kurzthaler2016intermediate, kurzthaler2017intermediate}}
First, in the large time and length scale, diffusion-like behavior is expected, as the persistent swimming is smeared out by frequent particle collisions (in the parameter region used in simulation, {$\tau_b/\tau_r \simeq 4\times 10^{-3}$}). 
Thus the correlation follows a near-exponential decay $\sim \exp(-D_{\text{eff}}q^2t)$ at $t\gg \tau_b$ and small $n_q$ (the dashed curve in Fig.~\ref{fig:sISF} (a)).
Whereas in MIPS at slow rotation $\Gamma \lesssim 1$, $F_s(q,t)$ relaxes in a much slower manner than the diffusive motion, reflecting particles frozen inside the stable dense droplet, confirming the particle exchange is an interfacial process in bulk phase separation.
Second, to observe the circular motion, one should look at large enough wavelength $\Lambda=L/n_q \gtrsim R$, thus we choose $n_q=1$.
Intriguingly, for fast rotation $\Gamma \gtrsim 1$, $F_s(q,t)$ oscillations around a finite plateau at the time scale around $\tau_w$, suggesting that the chiral swimming pattern of individual particle is partially preserved in the dynamical clustering state, even when dense liquid droplets are formed.
Finally, at large wave number $n_q=5$, the oscillations of $F_s(q,t)$ around zero within $\tau_r$ indicates a persistent motion $|\Delta \mathbf{r}(t)|=v_{\text{eff}}t$ at short wavelength, till at longer times the rotational diffusion washes out the oscillations.
The oscillation amplitude increases with $\Gamma$ indicating less velocity damping effect, and it finally saturates when further increase $\Gamma$ into a homogeneous state.
We notice that the instantaneous effective swimming velocity $v_e$ is independent of $\omega_0$, while the time averaged persistence  velocity $v_{\text{eff}}$ increases with $\omega_0$ as shown in the sISF.
This inconsistency suggests that fast enough rotation can ``relieve'' trapped particles from temporal fluctuations before being smeared out by the rotational diffusion, thus preserves the chiral swimming pattern and gives rise to the system spanning circulating current.

\section*{Conclusion and Discussion}
In conclusion, a  dynamical clustering state emerges in cABPs with sufficiently fast rotation and interrupts the conventional MIPS.
The underlying short wavelength instability originates from the combination of the conventional MIPS cohesion and the circulating current caused disintegration.
Different from most linear active particles, detailed balance cannot be restored at the continuum level for chiral active particles, which allows the non-zero curl current in steady states.
This is surprising, as there is no explicit alignment interaction between chiral active colloids to induce any collective spontaneous flow in the system~\cite{liebchen2017collective}.
More generally, the instability introduced by simply switching on monochromatic rotation might help design active colloids that can achieve complex spatio-temporal patterns formation.
Finally, our key finding of the well defined non-equilibrium steady state may also provide a platform to examine the generalization of effective thermodynamic concepts \cite{Solon_2018, PhysRevE.91.032117} previously developed in the context of MIPS.

\section*{acknowledgements}
This work has been supported by the Singapore Ministry of Education through the Academic Research Fund MOE2019-T2-2-010. We thank NSCC for granting computational resources.

\section*{Supplementary Material}
See the supplementary material for the derivation of the continuum theory and the linear stability analysis as well as a video for the dynamical clustering phase.

\section*{Data availability}
The data that support the findings of this study are available within the article and its supplementary material and from the corresponding author upon reasonable request.

\bibliography{references.bib}

\begin{thebibliography}{45}%
\makeatletter
\providecommand \@ifxundefined [1]{%
 \@ifx{#1\undefined}
}%
\providecommand \@ifnum [1]{%
 \ifnum #1\expandafter \@firstoftwo
 \else \expandafter \@secondoftwo
 \fi
}%
\providecommand \@ifx [1]{%
 \ifx #1\expandafter \@firstoftwo
 \else \expandafter \@secondoftwo
 \fi
}%
\providecommand \natexlab [1]{#1}%
\providecommand \enquote  [1]{``#1''}%
\providecommand \bibnamefont  [1]{#1}%
\providecommand \bibfnamefont [1]{#1}%
\providecommand \citenamefont [1]{#1}%
\providecommand \href@noop [0]{\@secondoftwo}%
\providecommand \href [0]{\begingroup \@sanitize@url \@href}%
\providecommand \@href[1]{\@@startlink{#1}\@@href}%
\providecommand \@@href[1]{\endgroup#1\@@endlink}%
\providecommand \@sanitize@url [0]{\catcode `\\12\catcode `\$12\catcode
  `\&12\catcode `\#12\catcode `\^12\catcode `\_12\catcode `\%12\relax}%
\providecommand \@@startlink[1]{}%
\providecommand \@@endlink[0]{}%
\providecommand \url  [0]{\begingroup\@sanitize@url \@url }%
\providecommand \@url [1]{\endgroup\@href {#1}{\urlprefix }}%
\providecommand \urlprefix  [0]{URL }%
\providecommand \Eprint [0]{\href }%
\providecommand \doibase [0]{https://doi.org/}%
\providecommand \selectlanguage [0]{\@gobble}%
\providecommand \bibinfo  [0]{\@secondoftwo}%
\providecommand \bibfield  [0]{\@secondoftwo}%
\providecommand \translation [1]{[#1]}%
\providecommand \BibitemOpen [0]{}%
\providecommand \bibitemStop [0]{}%
\providecommand \bibitemNoStop [0]{.\EOS\space}%
\providecommand \EOS [0]{\spacefactor3000\relax}%
\providecommand \BibitemShut  [1]{\csname bibitem#1\endcsname}%
\let\auto@bib@innerbib\@empty
\bibitem [{\citenamefont {van~der Waals}(1873)}]{vdwthesis}%
  \BibitemOpen
  \bibfield  {author} {\bibinfo {author} {\bibfnamefont {J.}~\bibnamefont
  {van~der Waals}},\ }\emph {\bibinfo {title} {Over de Continuiteit van den
  Gas- en Vloeistoftoestand}},\ \href@noop {} {Ph.D. thesis},\ \bibinfo
  {school} {Leiden University} (\bibinfo {year} {1873})\BibitemShut {NoStop}%
\bibitem [{\citenamefont {Fily}\ and\ \citenamefont
  {Marchetti}(2012)}]{PhysRevLett.108.235702}%
  \BibitemOpen
  \bibfield  {author} {\bibinfo {author} {\bibfnamefont {Y.}~\bibnamefont
  {Fily}}\ and\ \bibinfo {author} {\bibfnamefont {M.~C.}\ \bibnamefont
  {Marchetti}},\ }\bibfield  {title} {\bibinfo {title} {Athermal phase
  separation of self-propelled particles with no alignment},\ }\href
  {https://doi.org/10.1103/PhysRevLett.108.235702} {\bibfield  {journal}
  {\bibinfo  {journal} {Phys. Rev. Lett.}\ }\textbf {\bibinfo {volume} {108}},\
  \bibinfo {pages} {235702} (\bibinfo {year} {2012})}\BibitemShut {NoStop}%
\bibitem [{\citenamefont {Redner}\ \emph {et~al.}(2013)\citenamefont {Redner},
  \citenamefont {Hagan},\ and\ \citenamefont
  {Baskaran}}]{PhysRevLett.110.055701}%
  \BibitemOpen
  \bibfield  {author} {\bibinfo {author} {\bibfnamefont {G.~S.}\ \bibnamefont
  {Redner}}, \bibinfo {author} {\bibfnamefont {M.~F.}\ \bibnamefont {Hagan}},\
  and\ \bibinfo {author} {\bibfnamefont {A.}~\bibnamefont {Baskaran}},\
  }\bibfield  {title} {\bibinfo {title} {Structure and dynamics of a
  phase-separating active colloidal fluid},\ }\href
  {https://doi.org/10.1103/PhysRevLett.110.055701} {\bibfield  {journal}
  {\bibinfo  {journal} {Phys. Rev. Lett.}\ }\textbf {\bibinfo {volume} {110}},\
  \bibinfo {pages} {055701} (\bibinfo {year} {2013})}\BibitemShut {NoStop}%
\bibitem [{\citenamefont {Tailleur}\ and\ \citenamefont
  {Cates}(2008)}]{PhysRevLett.100.218103}%
  \BibitemOpen
  \bibfield  {author} {\bibinfo {author} {\bibfnamefont {J.}~\bibnamefont
  {Tailleur}}\ and\ \bibinfo {author} {\bibfnamefont {M.~E.}\ \bibnamefont
  {Cates}},\ }\bibfield  {title} {\bibinfo {title} {Statistical mechanics of
  interacting run-and-tumble bacteria},\ }\href
  {https://doi.org/10.1103/PhysRevLett.100.218103} {\bibfield  {journal}
  {\bibinfo  {journal} {Phys. Rev. Lett.}\ }\textbf {\bibinfo {volume} {100}},\
  \bibinfo {pages} {218103} (\bibinfo {year} {2008})}\BibitemShut {NoStop}%
\bibitem [{\citenamefont {Cates}\ and\ \citenamefont
  {Tailleur}(2013)}]{Cates_2013}%
  \BibitemOpen
  \bibfield  {author} {\bibinfo {author} {\bibfnamefont {M.~E.}\ \bibnamefont
  {Cates}}\ and\ \bibinfo {author} {\bibfnamefont {J.}~\bibnamefont
  {Tailleur}},\ }\bibfield  {title} {\bibinfo {title} {When are active brownian
  particles and run-and-tumble particles equivalent? consequences for
  motility-induced phase separation},\ }\href
  {https://doi.org/10.1209/0295-5075/101/20010} {\bibfield  {journal} {\bibinfo
   {journal} {Europhys. Lett.}\ }\textbf {\bibinfo {volume} {101}},\ \bibinfo
  {pages} {20010} (\bibinfo {year} {2013})}\BibitemShut {NoStop}%
\bibitem [{\citenamefont {Martin}\ \emph {et~al.}(2021)\citenamefont {Martin},
  \citenamefont {O'Byrne}, \citenamefont {Cates}, \citenamefont {Fodor},
  \citenamefont {Nardini}, \citenamefont {Tailleur},\ and\ \citenamefont {van
  Wijland}}]{PhysRevE.103.032607}%
  \BibitemOpen
  \bibfield  {author} {\bibinfo {author} {\bibfnamefont {D.}~\bibnamefont
  {Martin}}, \bibinfo {author} {\bibfnamefont {J.}~\bibnamefont {O'Byrne}},
  \bibinfo {author} {\bibfnamefont {M.~E.}\ \bibnamefont {Cates}}, \bibinfo
  {author} {\bibfnamefont {E.}~\bibnamefont {Fodor}}, \bibinfo {author}
  {\bibfnamefont {C.}~\bibnamefont {Nardini}}, \bibinfo {author} {\bibfnamefont
  {J.}~\bibnamefont {Tailleur}},\ and\ \bibinfo {author} {\bibfnamefont
  {F.}~\bibnamefont {van Wijland}},\ }\bibfield  {title} {\bibinfo {title}
  {Statistical mechanics of active ornstein-uhlenbeck particles},\ }\href
  {https://doi.org/10.1103/PhysRevE.103.032607} {\bibfield  {journal} {\bibinfo
   {journal} {Phys. Rev. E}\ }\textbf {\bibinfo {volume} {103}},\ \bibinfo
  {pages} {032607} (\bibinfo {year} {2021})}\BibitemShut {NoStop}%
\bibitem [{\citenamefont {Cates}\ and\ \citenamefont
  {Tailleur}(2015)}]{cates2015motility}%
  \BibitemOpen
  \bibfield  {author} {\bibinfo {author} {\bibfnamefont {M.~E.}\ \bibnamefont
  {Cates}}\ and\ \bibinfo {author} {\bibfnamefont {J.}~\bibnamefont
  {Tailleur}},\ }\bibfield  {title} {\bibinfo {title} {Motility-induced phase
  separation},\ }\href@noop {} {\bibfield  {journal} {\bibinfo  {journal}
  {Annu. Rev. Condens. Matter Phys.}\ }\textbf {\bibinfo {volume} {6}},\
  \bibinfo {pages} {219} (\bibinfo {year} {2015})}\BibitemShut {NoStop}%
\bibitem [{\citenamefont {Ma}\ \emph {et~al.}(2020)\citenamefont {Ma},
  \citenamefont {Yang},\ and\ \citenamefont {Ni}}]{ma2020}%
  \BibitemOpen
  \bibfield  {author} {\bibinfo {author} {\bibfnamefont {Z.}~\bibnamefont
  {Ma}}, \bibinfo {author} {\bibfnamefont {M.}~\bibnamefont {Yang}},\ and\
  \bibinfo {author} {\bibfnamefont {R.}~\bibnamefont {Ni}},\ }\bibfield
  {title} {\bibinfo {title} {Dynamic assembly of active colloids: Theory and
  simulation},\ }\href@noop {} {\bibfield  {journal} {\bibinfo  {journal} {Adv.
  Theory Simul.}\ }\textbf {\bibinfo {volume} {3}},\ \bibinfo {pages} {2000021}
  (\bibinfo {year} {2020})}\BibitemShut {NoStop}%
\bibitem [{\citenamefont {Solon}\ \emph {et~al.}(2018)\citenamefont {Solon},
  \citenamefont {Stenhammar}, \citenamefont {Cates}, \citenamefont {Kafri},\
  and\ \citenamefont {Tailleur}}]{Solon_2018}%
  \BibitemOpen
  \bibfield  {author} {\bibinfo {author} {\bibfnamefont {A.~P.}\ \bibnamefont
  {Solon}}, \bibinfo {author} {\bibfnamefont {J.}~\bibnamefont {Stenhammar}},
  \bibinfo {author} {\bibfnamefont {M.~E.}\ \bibnamefont {Cates}}, \bibinfo
  {author} {\bibfnamefont {Y.}~\bibnamefont {Kafri}},\ and\ \bibinfo {author}
  {\bibfnamefont {J.}~\bibnamefont {Tailleur}},\ }\bibfield  {title} {\bibinfo
  {title} {Generalized thermodynamics of motility-induced phase separation:
  phase equilibria, laplace pressure, and change of ensembles},\ }\href
  {https://doi.org/10.1088/1367-2630/aaccdd} {\bibfield  {journal} {\bibinfo
  {journal} {New J. Phys.}\ }\textbf {\bibinfo {volume} {20}},\ \bibinfo
  {pages} {075001} (\bibinfo {year} {2018})}\BibitemShut {NoStop}%
\bibitem [{\citenamefont {B{\"o}hmer}\ \emph {et~al.}(2005)\citenamefont
  {B{\"o}hmer}, \citenamefont {Van}, \citenamefont {Weyand}, \citenamefont
  {Hagen}, \citenamefont {Beyermann}, \citenamefont {Matsumoto}, \citenamefont
  {Hoshi}, \citenamefont {Hildebrand},\ and\ \citenamefont
  {Kaupp}}]{bohmer2005ca2+}%
  \BibitemOpen
  \bibfield  {author} {\bibinfo {author} {\bibfnamefont {M.}~\bibnamefont
  {B{\"o}hmer}}, \bibinfo {author} {\bibfnamefont {Q.}~\bibnamefont {Van}},
  \bibinfo {author} {\bibfnamefont {I.}~\bibnamefont {Weyand}}, \bibinfo
  {author} {\bibfnamefont {V.}~\bibnamefont {Hagen}}, \bibinfo {author}
  {\bibfnamefont {M.}~\bibnamefont {Beyermann}}, \bibinfo {author}
  {\bibfnamefont {M.}~\bibnamefont {Matsumoto}}, \bibinfo {author}
  {\bibfnamefont {M.}~\bibnamefont {Hoshi}}, \bibinfo {author} {\bibfnamefont
  {E.}~\bibnamefont {Hildebrand}},\ and\ \bibinfo {author} {\bibfnamefont
  {U.~B.}\ \bibnamefont {Kaupp}},\ }\bibfield  {title} {\bibinfo {title} {Ca2+
  spikes in the flagellum control chemotactic behavior of sperm},\ }\href@noop
  {} {\bibfield  {journal} {\bibinfo  {journal} {EMBO J.}\ }\textbf {\bibinfo
  {volume} {24}},\ \bibinfo {pages} {2741} (\bibinfo {year}
  {2005})}\BibitemShut {NoStop}%
\bibitem [{\citenamefont {Taktikos}\ \emph {et~al.}(2011)\citenamefont
  {Taktikos}, \citenamefont {Zaburdaev},\ and\ \citenamefont
  {Stark}}]{taktikos2011modeling}%
  \BibitemOpen
  \bibfield  {author} {\bibinfo {author} {\bibfnamefont {J.}~\bibnamefont
  {Taktikos}}, \bibinfo {author} {\bibfnamefont {V.}~\bibnamefont
  {Zaburdaev}},\ and\ \bibinfo {author} {\bibfnamefont {H.}~\bibnamefont
  {Stark}},\ }\bibfield  {title} {\bibinfo {title} {Modeling a self-propelled
  autochemotactic walker},\ }\href {https://doi.org/10.1103/PhysRevE.84.041924}
  {\bibfield  {journal} {\bibinfo  {journal} {Phys. Rev. E}\ }\textbf {\bibinfo
  {volume} {84}},\ \bibinfo {pages} {041924} (\bibinfo {year}
  {2011})}\BibitemShut {NoStop}%
\bibitem [{\citenamefont {DiLuzio}\ \emph {et~al.}(2005)\citenamefont
  {DiLuzio}, \citenamefont {Turner}, \citenamefont {Mayer}, \citenamefont
  {Garstecki}, \citenamefont {Weibel}, \citenamefont {Berg},\ and\
  \citenamefont {Whitesides}}]{diluzio2005escherichia}%
  \BibitemOpen
  \bibfield  {author} {\bibinfo {author} {\bibfnamefont {W.~R.}\ \bibnamefont
  {DiLuzio}}, \bibinfo {author} {\bibfnamefont {L.}~\bibnamefont {Turner}},
  \bibinfo {author} {\bibfnamefont {M.}~\bibnamefont {Mayer}}, \bibinfo
  {author} {\bibfnamefont {P.}~\bibnamefont {Garstecki}}, \bibinfo {author}
  {\bibfnamefont {D.~B.}\ \bibnamefont {Weibel}}, \bibinfo {author}
  {\bibfnamefont {H.~C.}\ \bibnamefont {Berg}},\ and\ \bibinfo {author}
  {\bibfnamefont {G.~M.}\ \bibnamefont {Whitesides}},\ }\bibfield  {title}
  {\bibinfo {title} {Escherichia coli swim on the right-hand side},\
  }\href@noop {} {\bibfield  {journal} {\bibinfo  {journal} {Nature}\ }\textbf
  {\bibinfo {volume} {435}},\ \bibinfo {pages} {1271} (\bibinfo {year}
  {2005})}\BibitemShut {NoStop}%
\bibitem [{\citenamefont {Di~Leonardo}\ \emph {et~al.}(2011)\citenamefont
  {Di~Leonardo}, \citenamefont {Dell'Arciprete}, \citenamefont {Angelani},\
  and\ \citenamefont {Iebba}}]{di2011swimming}%
  \BibitemOpen
  \bibfield  {author} {\bibinfo {author} {\bibfnamefont {R.}~\bibnamefont
  {Di~Leonardo}}, \bibinfo {author} {\bibfnamefont {D.}~\bibnamefont
  {Dell'Arciprete}}, \bibinfo {author} {\bibfnamefont {L.}~\bibnamefont
  {Angelani}},\ and\ \bibinfo {author} {\bibfnamefont {V.}~\bibnamefont
  {Iebba}},\ }\bibfield  {title} {\bibinfo {title} {Swimming with an image},\
  }\href {https://doi.org/10.1103/PhysRevLett.106.038101} {\bibfield  {journal}
  {\bibinfo  {journal} {Phys. Rev. Lett.}\ }\textbf {\bibinfo {volume} {106}},\
  \bibinfo {pages} {038101} (\bibinfo {year} {2011})}\BibitemShut {NoStop}%
\bibitem [{\citenamefont {Martinez}\ \emph {et~al.}(2012)\citenamefont
  {Martinez}, \citenamefont {Besseling}, \citenamefont {Croze}, \citenamefont
  {Tailleur}, \citenamefont {Reufer}, \citenamefont {Schwarz-Linek},
  \citenamefont {Wilson}, \citenamefont {Bees},\ and\ \citenamefont
  {Poon}}]{martinez2012differential}%
  \BibitemOpen
  \bibfield  {author} {\bibinfo {author} {\bibfnamefont {V.~A.}\ \bibnamefont
  {Martinez}}, \bibinfo {author} {\bibfnamefont {R.}~\bibnamefont {Besseling}},
  \bibinfo {author} {\bibfnamefont {O.~A.}\ \bibnamefont {Croze}}, \bibinfo
  {author} {\bibfnamefont {J.}~\bibnamefont {Tailleur}}, \bibinfo {author}
  {\bibfnamefont {M.}~\bibnamefont {Reufer}}, \bibinfo {author} {\bibfnamefont
  {J.}~\bibnamefont {Schwarz-Linek}}, \bibinfo {author} {\bibfnamefont {L.~G.}\
  \bibnamefont {Wilson}}, \bibinfo {author} {\bibfnamefont {M.~A.}\
  \bibnamefont {Bees}},\ and\ \bibinfo {author} {\bibfnamefont {W.~C.}\
  \bibnamefont {Poon}},\ }\bibfield  {title} {\bibinfo {title} {Differential
  dynamic microscopy: A high-throughput method for characterizing the motility
  of microorganisms},\ }\href@noop {} {\bibfield  {journal} {\bibinfo
  {journal} {Biophys. J.}\ }\textbf {\bibinfo {volume} {103}},\ \bibinfo
  {pages} {1637} (\bibinfo {year} {2012})}\BibitemShut {NoStop}%
\bibitem [{\citenamefont {Utada}\ \emph {et~al.}(2014)\citenamefont {Utada},
  \citenamefont {Bennett}, \citenamefont {Fong}, \citenamefont {Gibiansky},
  \citenamefont {Yildiz}, \citenamefont {Golestanian},\ and\ \citenamefont
  {Wong}}]{utada2014vibrio}%
  \BibitemOpen
  \bibfield  {author} {\bibinfo {author} {\bibfnamefont {A.~S.}\ \bibnamefont
  {Utada}}, \bibinfo {author} {\bibfnamefont {R.~R.}\ \bibnamefont {Bennett}},
  \bibinfo {author} {\bibfnamefont {J.~C.}\ \bibnamefont {Fong}}, \bibinfo
  {author} {\bibfnamefont {M.~L.}\ \bibnamefont {Gibiansky}}, \bibinfo {author}
  {\bibfnamefont {F.~H.}\ \bibnamefont {Yildiz}}, \bibinfo {author}
  {\bibfnamefont {R.}~\bibnamefont {Golestanian}},\ and\ \bibinfo {author}
  {\bibfnamefont {G.~C.}\ \bibnamefont {Wong}},\ }\bibfield  {title} {\bibinfo
  {title} {Vibrio cholerae use pili and flagella synergistically to effect
  motility switching and conditional surface attachment},\ }\href@noop {}
  {\bibfield  {journal} {\bibinfo  {journal} {Nat. Commun.}\ }\textbf {\bibinfo
  {volume} {5}},\ \bibinfo {pages} {4913} (\bibinfo {year} {2014})}\BibitemShut
  {NoStop}%
\bibitem [{\citenamefont {Gibbs}\ and\ \citenamefont
  {Zhao}(2009)}]{gibbs2009design}%
  \BibitemOpen
  \bibfield  {author} {\bibinfo {author} {\bibfnamefont {J.}~\bibnamefont
  {Gibbs}}\ and\ \bibinfo {author} {\bibfnamefont {Y.-P.}\ \bibnamefont
  {Zhao}},\ }\bibfield  {title} {\bibinfo {title} {Design and characterization
  of rotational multicomponent catalytic nanomotors},\ }\href@noop {}
  {\bibfield  {journal} {\bibinfo  {journal} {Small}\ }\textbf {\bibinfo
  {volume} {5}},\ \bibinfo {pages} {2304} (\bibinfo {year} {2009})}\BibitemShut
  {NoStop}%
\bibitem [{\citenamefont {Gibbs}\ \emph {et~al.}(2011)\citenamefont {Gibbs},
  \citenamefont {Kothari}, \citenamefont {Saintillan},\ and\ \citenamefont
  {Zhao}}]{gibbs2011geometrically}%
  \BibitemOpen
  \bibfield  {author} {\bibinfo {author} {\bibfnamefont {J.}~\bibnamefont
  {Gibbs}}, \bibinfo {author} {\bibfnamefont {S.}~\bibnamefont {Kothari}},
  \bibinfo {author} {\bibfnamefont {D.}~\bibnamefont {Saintillan}},\ and\
  \bibinfo {author} {\bibfnamefont {Y.-P.}\ \bibnamefont {Zhao}},\ }\bibfield
  {title} {\bibinfo {title} {Geometrically designing the kinematic behavior of
  catalytic nanomotors},\ }\href@noop {} {\bibfield  {journal} {\bibinfo
  {journal} {Nano Lett.}\ }\textbf {\bibinfo {volume} {11}},\ \bibinfo {pages}
  {2543} (\bibinfo {year} {2011})}\BibitemShut {NoStop}%
\bibitem [{\citenamefont {K\"ummel}\ \emph {et~al.}(2013)\citenamefont
  {K\"ummel}, \citenamefont {ten Hagen}, \citenamefont {Wittkowski},
  \citenamefont {Buttinoni}, \citenamefont {Eichhorn}, \citenamefont {Volpe},
  \citenamefont {L\"owen},\ and\ \citenamefont
  {Bechinger}}]{kummel2013circular}%
  \BibitemOpen
  \bibfield  {author} {\bibinfo {author} {\bibfnamefont {F.}~\bibnamefont
  {K\"ummel}}, \bibinfo {author} {\bibfnamefont {B.}~\bibnamefont {ten Hagen}},
  \bibinfo {author} {\bibfnamefont {R.}~\bibnamefont {Wittkowski}}, \bibinfo
  {author} {\bibfnamefont {I.}~\bibnamefont {Buttinoni}}, \bibinfo {author}
  {\bibfnamefont {R.}~\bibnamefont {Eichhorn}}, \bibinfo {author}
  {\bibfnamefont {G.}~\bibnamefont {Volpe}}, \bibinfo {author} {\bibfnamefont
  {H.}~\bibnamefont {L\"owen}},\ and\ \bibinfo {author} {\bibfnamefont
  {C.}~\bibnamefont {Bechinger}},\ }\bibfield  {title} {\bibinfo {title}
  {Circular motion of asymmetric self-propelling particles},\ }\href
  {https://doi.org/10.1103/PhysRevLett.110.198302} {\bibfield  {journal}
  {\bibinfo  {journal} {Phys. Rev. Lett.}\ }\textbf {\bibinfo {volume} {110}},\
  \bibinfo {pages} {198302} (\bibinfo {year} {2013})}\BibitemShut {NoStop}%
\bibitem [{\citenamefont {Campbell}\ \emph {et~al.}(2017)\citenamefont
  {Campbell}, \citenamefont {Wittkowski}, \citenamefont {Ten~Hagen},
  \citenamefont {L{\"o}wen},\ and\ \citenamefont
  {Ebbens}}]{campbell2017helical}%
  \BibitemOpen
  \bibfield  {author} {\bibinfo {author} {\bibfnamefont {A.~I.}\ \bibnamefont
  {Campbell}}, \bibinfo {author} {\bibfnamefont {R.}~\bibnamefont
  {Wittkowski}}, \bibinfo {author} {\bibfnamefont {B.}~\bibnamefont
  {Ten~Hagen}}, \bibinfo {author} {\bibfnamefont {H.}~\bibnamefont
  {L{\"o}wen}},\ and\ \bibinfo {author} {\bibfnamefont {S.~J.}\ \bibnamefont
  {Ebbens}},\ }\bibfield  {title} {\bibinfo {title} {Helical paths, gravitaxis,
  and separation phenomena for mass-anisotropic self-propelling colloids:
  Experiment versus theory},\ }\href@noop {} {\bibfield  {journal} {\bibinfo
  {journal} {J. Chem. Phys.}\ }\textbf {\bibinfo {volume} {147}},\ \bibinfo
  {pages} {084905} (\bibinfo {year} {2017})}\BibitemShut {NoStop}%
\bibitem [{\citenamefont {Archer}\ \emph {et~al.}(2015)\citenamefont {Archer},
  \citenamefont {Campbell},\ and\ \citenamefont {Ebbens}}]{archer2015glancing}%
  \BibitemOpen
  \bibfield  {author} {\bibinfo {author} {\bibfnamefont {R.}~\bibnamefont
  {Archer}}, \bibinfo {author} {\bibfnamefont {A.}~\bibnamefont {Campbell}},\
  and\ \bibinfo {author} {\bibfnamefont {S.}~\bibnamefont {Ebbens}},\
  }\bibfield  {title} {\bibinfo {title} {Glancing angle metal evaporation
  synthesis of catalytic swimming janus colloids with well defined angular
  velocity},\ }\href@noop {} {\bibfield  {journal} {\bibinfo  {journal} {Soft
  Matter}\ }\textbf {\bibinfo {volume} {11}},\ \bibinfo {pages} {6872}
  (\bibinfo {year} {2015})}\BibitemShut {NoStop}%
\bibitem [{\citenamefont {Ten~Hagen}\ \emph {et~al.}(2014)\citenamefont
  {Ten~Hagen}, \citenamefont {K{\"u}mmel}, \citenamefont {Wittkowski},
  \citenamefont {Takagi}, \citenamefont {L{\"o}wen},\ and\ \citenamefont
  {Bechinger}}]{ten2014gravitaxis}%
  \BibitemOpen
  \bibfield  {author} {\bibinfo {author} {\bibfnamefont {B.}~\bibnamefont
  {Ten~Hagen}}, \bibinfo {author} {\bibfnamefont {F.}~\bibnamefont
  {K{\"u}mmel}}, \bibinfo {author} {\bibfnamefont {R.}~\bibnamefont
  {Wittkowski}}, \bibinfo {author} {\bibfnamefont {D.}~\bibnamefont {Takagi}},
  \bibinfo {author} {\bibfnamefont {H.}~\bibnamefont {L{\"o}wen}},\ and\
  \bibinfo {author} {\bibfnamefont {C.}~\bibnamefont {Bechinger}},\ }\bibfield
  {title} {\bibinfo {title} {Gravitaxis of asymmetric self-propelled colloidal
  particles},\ }\href@noop {} {\bibfield  {journal} {\bibinfo  {journal} {Nat.
  Commun.}\ }\textbf {\bibinfo {volume} {5}},\ \bibinfo {pages} {4829}
  (\bibinfo {year} {2014})}\BibitemShut {NoStop}%
\bibitem [{\citenamefont {Mano}\ \emph {et~al.}(2017)\citenamefont {Mano},
  \citenamefont {Delfau}, \citenamefont {Iwasawa},\ and\ \citenamefont
  {Sano}}]{ManoE2580}%
  \BibitemOpen
  \bibfield  {author} {\bibinfo {author} {\bibfnamefont {T.}~\bibnamefont
  {Mano}}, \bibinfo {author} {\bibfnamefont {J.-B.}\ \bibnamefont {Delfau}},
  \bibinfo {author} {\bibfnamefont {J.}~\bibnamefont {Iwasawa}},\ and\ \bibinfo
  {author} {\bibfnamefont {M.}~\bibnamefont {Sano}},\ }\bibfield  {title}
  {\bibinfo {title} {Optimal run-and-tumble{\textendash}based transportation of
  a janus particle with active steering},\ }\href
  {https://doi.org/10.1073/pnas.1616013114} {\bibfield  {journal} {\bibinfo
  {journal} {Proc. Nat. Acad. Sci. USA}\ }\textbf {\bibinfo {volume} {114}},\
  \bibinfo {pages} {E2580} (\bibinfo {year} {2017})}\BibitemShut {NoStop}%
\bibitem [{\citenamefont {Chepizhko}\ and\ \citenamefont
  {Franosch}(2019)}]{C8SM02030B}%
  \BibitemOpen
  \bibfield  {author} {\bibinfo {author} {\bibfnamefont {O.}~\bibnamefont
  {Chepizhko}}\ and\ \bibinfo {author} {\bibfnamefont {T.}~\bibnamefont
  {Franosch}},\ }\bibfield  {title} {\bibinfo {title} {Ideal circle
  microswimmers in crowded media},\ }\href {https://doi.org/10.1039/C8SM02030B}
  {\bibfield  {journal} {\bibinfo  {journal} {Soft Matter}\ }\textbf {\bibinfo
  {volume} {15}},\ \bibinfo {pages} {452} (\bibinfo {year} {2019})}\BibitemShut
  {NoStop}%
\bibitem [{\citenamefont {Bechinger}\ \emph {et~al.}(2016)\citenamefont
  {Bechinger}, \citenamefont {Di~Leonardo}, \citenamefont {L\"owen},
  \citenamefont {Reichhardt}, \citenamefont {Volpe},\ and\ \citenamefont
  {Volpe}}]{RevModPhys2016}%
  \BibitemOpen
  \bibfield  {author} {\bibinfo {author} {\bibfnamefont {C.}~\bibnamefont
  {Bechinger}}, \bibinfo {author} {\bibfnamefont {R.}~\bibnamefont
  {Di~Leonardo}}, \bibinfo {author} {\bibfnamefont {H.}~\bibnamefont
  {L\"owen}}, \bibinfo {author} {\bibfnamefont {C.}~\bibnamefont {Reichhardt}},
  \bibinfo {author} {\bibfnamefont {G.}~\bibnamefont {Volpe}},\ and\ \bibinfo
  {author} {\bibfnamefont {G.}~\bibnamefont {Volpe}},\ }\bibfield  {title}
  {\bibinfo {title} {Active particles in complex and crowded environments},\
  }\href {https://doi.org/10.1103/RevModPhys.88.045006} {\bibfield  {journal}
  {\bibinfo  {journal} {Rev. Mod. Phys.}\ }\textbf {\bibinfo {volume} {88}},\
  \bibinfo {pages} {045006} (\bibinfo {year} {2016})}\BibitemShut {NoStop}%
\bibitem [{\citenamefont {Liebchen}\ and\ \citenamefont
  {Levis}(2017)}]{liebchen2017collective}%
  \BibitemOpen
  \bibfield  {author} {\bibinfo {author} {\bibfnamefont {B.}~\bibnamefont
  {Liebchen}}\ and\ \bibinfo {author} {\bibfnamefont {D.}~\bibnamefont
  {Levis}},\ }\bibfield  {title} {\bibinfo {title} {Collective behavior of
  chiral active matter: Pattern formation and enhanced flocking},\ }\href
  {https://doi.org/10.1103/PhysRevLett.119.058002} {\bibfield  {journal}
  {\bibinfo  {journal} {Phys. Rev. Lett.}\ }\textbf {\bibinfo {volume} {119}},\
  \bibinfo {pages} {058002} (\bibinfo {year} {2017})}\BibitemShut {NoStop}%
\bibitem [{\citenamefont {Levis}\ and\ \citenamefont
  {Liebchen}(2018)}]{levis2018micro}%
  \BibitemOpen
  \bibfield  {author} {\bibinfo {author} {\bibfnamefont {D.}~\bibnamefont
  {Levis}}\ and\ \bibinfo {author} {\bibfnamefont {B.}~\bibnamefont
  {Liebchen}},\ }\bibfield  {title} {\bibinfo {title} {Micro-flock patterns and
  macro-clusters in chiral active brownian disks},\ }\href@noop {} {\bibfield
  {journal} {\bibinfo  {journal} {J. Phys. Condens. Matter}\ }\textbf {\bibinfo
  {volume} {30}},\ \bibinfo {pages} {084001} (\bibinfo {year}
  {2018})}\BibitemShut {NoStop}%
\bibitem [{\citenamefont {Liao}\ and\ \citenamefont
  {Klapp}(2018)}]{liao2018clustering}%
  \BibitemOpen
  \bibfield  {author} {\bibinfo {author} {\bibfnamefont {G.-J.}\ \bibnamefont
  {Liao}}\ and\ \bibinfo {author} {\bibfnamefont {S.~H.}\ \bibnamefont
  {Klapp}},\ }\bibfield  {title} {\bibinfo {title} {Clustering and phase
  separation of circle swimmers dispersed in a monolayer},\ }\href@noop {}
  {\bibfield  {journal} {\bibinfo  {journal} {Soft matter}\ }\textbf {\bibinfo
  {volume} {14}},\ \bibinfo {pages} {7873} (\bibinfo {year}
  {2018})}\BibitemShut {NoStop}%
\bibitem [{\citenamefont {Kaiser}\ and\ \citenamefont
  {L\"owen}(2013)}]{PhysRevE.87.032712}%
  \BibitemOpen
  \bibfield  {author} {\bibinfo {author} {\bibfnamefont {A.}~\bibnamefont
  {Kaiser}}\ and\ \bibinfo {author} {\bibfnamefont {H.}~\bibnamefont
  {L\"owen}},\ }\bibfield  {title} {\bibinfo {title} {Vortex arrays as emergent
  collective phenomena for circle swimmers},\ }\href
  {https://doi.org/10.1103/PhysRevE.87.032712} {\bibfield  {journal} {\bibinfo
  {journal} {Phys. Rev. E}\ }\textbf {\bibinfo {volume} {87}},\ \bibinfo
  {pages} {032712} (\bibinfo {year} {2013})}\BibitemShut {NoStop}%
\bibitem [{\citenamefont {Liu}\ \emph {et~al.}(2019)\citenamefont {Liu},
  \citenamefont {Yang}, \citenamefont {Li},\ and\ \citenamefont
  {Feng}}]{liu2019collective}%
  \BibitemOpen
  \bibfield  {author} {\bibinfo {author} {\bibfnamefont {Y.}~\bibnamefont
  {Liu}}, \bibinfo {author} {\bibfnamefont {Y.}~\bibnamefont {Yang}}, \bibinfo
  {author} {\bibfnamefont {B.}~\bibnamefont {Li}},\ and\ \bibinfo {author}
  {\bibfnamefont {X.-Q.}\ \bibnamefont {Feng}},\ }\bibfield  {title} {\bibinfo
  {title} {Collective oscillation in dense suspension of self-propelled chiral
  rods},\ }\href@noop {} {\bibfield  {journal} {\bibinfo  {journal} {Soft
  matter}\ }\textbf {\bibinfo {volume} {15}},\ \bibinfo {pages} {2999}
  (\bibinfo {year} {2019})}\BibitemShut {NoStop}%
\bibitem [{\citenamefont {Bickmann}\ \emph {et~al.}(2020)\citenamefont
  {Bickmann}, \citenamefont {Br{\"o}ker}, \citenamefont {Jeggle},\ and\
  \citenamefont {Wittkowski}}]{bickmann2020analytical}%
  \BibitemOpen
  \bibfield  {author} {\bibinfo {author} {\bibfnamefont {J.}~\bibnamefont
  {Bickmann}}, \bibinfo {author} {\bibfnamefont {S.}~\bibnamefont
  {Br{\"o}ker}}, \bibinfo {author} {\bibfnamefont {J.}~\bibnamefont {Jeggle}},\
  and\ \bibinfo {author} {\bibfnamefont {R.}~\bibnamefont {Wittkowski}},\
  }\bibfield  {title} {\bibinfo {title} {Analytical approach to chiral active
  systems: suppressed phase separation of interacting brownian circle
  swimmers},\ }\href@noop {} {\bibfield  {journal} {\bibinfo  {journal} {arXiv
  preprint arXiv:2010.05262}\ } (\bibinfo {year} {2020})}\BibitemShut {NoStop}%
\bibitem [{\citenamefont {Lei}\ \emph {et~al.}(2019)\citenamefont {Lei},
  \citenamefont {Ciamarra},\ and\ \citenamefont {Ni}}]{lei2019nonequilibrium}%
  \BibitemOpen
  \bibfield  {author} {\bibinfo {author} {\bibfnamefont {Q.-L.}\ \bibnamefont
  {Lei}}, \bibinfo {author} {\bibfnamefont {M.~P.}\ \bibnamefont {Ciamarra}},\
  and\ \bibinfo {author} {\bibfnamefont {R.}~\bibnamefont {Ni}},\ }\bibfield
  {title} {\bibinfo {title} {Nonequilibrium strongly hyperuniform fluids of
  circle active particles with large local density fluctuations},\ }\href@noop
  {} {\bibfield  {journal} {\bibinfo  {journal} {Sci. Adv.}\ }\textbf {\bibinfo
  {volume} {5}},\ \bibinfo {pages} {eaau7423} (\bibinfo {year}
  {2019})}\BibitemShut {NoStop}%
\bibitem [{\citenamefont {Lei}\ and\ \citenamefont
  {Ni}(2019)}]{lei2019nonequilibrium2}%
  \BibitemOpen
  \bibfield  {author} {\bibinfo {author} {\bibfnamefont {Q.-L.}\ \bibnamefont
  {Lei}}\ and\ \bibinfo {author} {\bibfnamefont {R.}~\bibnamefont {Ni}},\
  }\bibfield  {title} {\bibinfo {title} {Hydrodynamics of random-organizing
  hyperuniform fluids},\ }\href@noop {} {\bibfield  {journal} {\bibinfo
  {journal} {Proc. Natl Acad. Sci. USA,}\ }\textbf {\bibinfo {volume} {116}},\
  \bibinfo {pages} {22983} (\bibinfo {year} {2019})}\BibitemShut {NoStop}%
\bibitem [{\citenamefont {Theurkauff}\ \emph {et~al.}(2012)\citenamefont
  {Theurkauff}, \citenamefont {Cottin-Bizonne}, \citenamefont {Palacci},
  \citenamefont {Ybert},\ and\ \citenamefont
  {Bocquet}}]{PhysRevLett.108.268303}%
  \BibitemOpen
  \bibfield  {author} {\bibinfo {author} {\bibfnamefont {I.}~\bibnamefont
  {Theurkauff}}, \bibinfo {author} {\bibfnamefont {C.}~\bibnamefont
  {Cottin-Bizonne}}, \bibinfo {author} {\bibfnamefont {J.}~\bibnamefont
  {Palacci}}, \bibinfo {author} {\bibfnamefont {C.}~\bibnamefont {Ybert}},\
  and\ \bibinfo {author} {\bibfnamefont {L.}~\bibnamefont {Bocquet}},\
  }\bibfield  {title} {\bibinfo {title} {Dynamic clustering in active colloidal
  suspensions with chemical signaling},\ }\href
  {https://doi.org/10.1103/PhysRevLett.108.268303} {\bibfield  {journal}
  {\bibinfo  {journal} {Phys. Rev. Lett.}\ }\textbf {\bibinfo {volume} {108}},\
  \bibinfo {pages} {268303} (\bibinfo {year} {2012})}\BibitemShut {NoStop}%
\bibitem [{\citenamefont {Liebchen}\ \emph {et~al.}(2015)\citenamefont
  {Liebchen}, \citenamefont {Marenduzzo}, \citenamefont {Pagonabarraga},\ and\
  \citenamefont {Cates}}]{PhysRevLett.115.258301}%
  \BibitemOpen
  \bibfield  {author} {\bibinfo {author} {\bibfnamefont {B.}~\bibnamefont
  {Liebchen}}, \bibinfo {author} {\bibfnamefont {D.}~\bibnamefont
  {Marenduzzo}}, \bibinfo {author} {\bibfnamefont {I.}~\bibnamefont
  {Pagonabarraga}},\ and\ \bibinfo {author} {\bibfnamefont {M.~E.}\
  \bibnamefont {Cates}},\ }\bibfield  {title} {\bibinfo {title} {Clustering and
  pattern formation in chemorepulsive active colloids},\ }\href
  {https://doi.org/10.1103/PhysRevLett.115.258301} {\bibfield  {journal}
  {\bibinfo  {journal} {Phys. Rev. Lett.}\ }\textbf {\bibinfo {volume} {115}},\
  \bibinfo {pages} {258301} (\bibinfo {year} {2015})}\BibitemShut {NoStop}%
\bibitem [{\citenamefont {Ma}\ \emph {et~al.}(2017)\citenamefont {Ma},
  \citenamefont {Lei},\ and\ \citenamefont {Ni}}]{ma2017driving}%
  \BibitemOpen
  \bibfield  {author} {\bibinfo {author} {\bibfnamefont {Z.}~\bibnamefont
  {Ma}}, \bibinfo {author} {\bibfnamefont {Q.-L.}\ \bibnamefont {Lei}},\ and\
  \bibinfo {author} {\bibfnamefont {R.}~\bibnamefont {Ni}},\ }\bibfield
  {title} {\bibinfo {title} {Driving dynamic colloidal assembly using eccentric
  self-propelled colloids},\ }\href@noop {} {\bibfield  {journal} {\bibinfo
  {journal} {Soft Matter}\ }\textbf {\bibinfo {volume} {13}},\ \bibinfo {pages}
  {8940} (\bibinfo {year} {2017})}\BibitemShut {NoStop}%
\bibitem [{\citenamefont {Bialk{\'e}}\ \emph {et~al.}(2013)\citenamefont
  {Bialk{\'e}}, \citenamefont {L{\"o}wen},\ and\ \citenamefont
  {Speck}}]{bialke2013microscopic}%
  \BibitemOpen
  \bibfield  {author} {\bibinfo {author} {\bibfnamefont {J.}~\bibnamefont
  {Bialk{\'e}}}, \bibinfo {author} {\bibfnamefont {H.}~\bibnamefont
  {L{\"o}wen}},\ and\ \bibinfo {author} {\bibfnamefont {T.}~\bibnamefont
  {Speck}},\ }\bibfield  {title} {\bibinfo {title} {Microscopic theory for the
  phase separation of self-propelled repulsive disks},\ }\href@noop {}
  {\bibfield  {journal} {\bibinfo  {journal} {Europhys. Lett.}\ }\textbf
  {\bibinfo {volume} {103}},\ \bibinfo {pages} {30008} (\bibinfo {year}
  {2013})}\BibitemShut {NoStop}%
\bibitem [{\citenamefont {Speck}\ \emph {et~al.}(2014)\citenamefont {Speck},
  \citenamefont {Bialk\'e}, \citenamefont {Menzel},\ and\ \citenamefont
  {L\"owen}}]{speck2014effective}%
  \BibitemOpen
  \bibfield  {author} {\bibinfo {author} {\bibfnamefont {T.}~\bibnamefont
  {Speck}}, \bibinfo {author} {\bibfnamefont {J.}~\bibnamefont {Bialk\'e}},
  \bibinfo {author} {\bibfnamefont {A.~M.}\ \bibnamefont {Menzel}},\ and\
  \bibinfo {author} {\bibfnamefont {H.}~\bibnamefont {L\"owen}},\ }\bibfield
  {title} {\bibinfo {title} {Effective cahn-hilliard equation for the phase
  separation of active brownian particles},\ }\href
  {https://doi.org/10.1103/PhysRevLett.112.218304} {\bibfield  {journal}
  {\bibinfo  {journal} {Phys. Rev. Lett.}\ }\textbf {\bibinfo {volume} {112}},\
  \bibinfo {pages} {218304} (\bibinfo {year} {2014})}\BibitemShut {NoStop}%
\bibitem [{\citenamefont {Speck}\ \emph {et~al.}(2015)\citenamefont {Speck},
  \citenamefont {Menzel}, \citenamefont {Bialk{\'e}},\ and\ \citenamefont
  {L{\"o}wen}}]{speck2015dynamical}%
  \BibitemOpen
  \bibfield  {author} {\bibinfo {author} {\bibfnamefont {T.}~\bibnamefont
  {Speck}}, \bibinfo {author} {\bibfnamefont {A.~M.}\ \bibnamefont {Menzel}},
  \bibinfo {author} {\bibfnamefont {J.}~\bibnamefont {Bialk{\'e}}},\ and\
  \bibinfo {author} {\bibfnamefont {H.}~\bibnamefont {L{\"o}wen}},\ }\bibfield
  {title} {\bibinfo {title} {Dynamical mean-field theory and weakly non-linear
  analysis for the phase separation of active brownian particles},\ }\href@noop
  {} {\bibfield  {journal} {\bibinfo  {journal} {J. Chem. Phys.}\ }\textbf
  {\bibinfo {volume} {142}},\ \bibinfo {pages} {224109} (\bibinfo {year}
  {2015})}\BibitemShut {NoStop}%
\bibitem [{\citenamefont {Caporusso}\ \emph {et~al.}(2020)\citenamefont
  {Caporusso}, \citenamefont {Digregorio}, \citenamefont {Levis}, \citenamefont
  {Cugliandolo},\ and\ \citenamefont {Gonnella}}]{PhysRevLett.125.178004}%
  \BibitemOpen
  \bibfield  {author} {\bibinfo {author} {\bibfnamefont {C.~B.}\ \bibnamefont
  {Caporusso}}, \bibinfo {author} {\bibfnamefont {P.}~\bibnamefont
  {Digregorio}}, \bibinfo {author} {\bibfnamefont {D.}~\bibnamefont {Levis}},
  \bibinfo {author} {\bibfnamefont {L.~F.}\ \bibnamefont {Cugliandolo}},\ and\
  \bibinfo {author} {\bibfnamefont {G.}~\bibnamefont {Gonnella}},\ }\bibfield
  {title} {\bibinfo {title} {Motility-induced microphase and macrophase
  separation in a two-dimensional active brownian particle system},\ }\href
  {https://doi.org/10.1103/PhysRevLett.125.178004} {\bibfield  {journal}
  {\bibinfo  {journal} {Phys. Rev. Lett.}\ }\textbf {\bibinfo {volume} {125}},\
  \bibinfo {pages} {178004} (\bibinfo {year} {2020})}\BibitemShut {NoStop}%
\bibitem [{\citenamefont {Shi}\ \emph {et~al.}(2020)\citenamefont {Shi},
  \citenamefont {Fausti}, \citenamefont {Chat\'e}, \citenamefont {Nardini},\
  and\ \citenamefont {Solon}}]{PhysRevLett.125.168001}%
  \BibitemOpen
  \bibfield  {author} {\bibinfo {author} {\bibfnamefont {X.-Q.}\ \bibnamefont
  {Shi}}, \bibinfo {author} {\bibfnamefont {G.}~\bibnamefont {Fausti}},
  \bibinfo {author} {\bibfnamefont {H.}~\bibnamefont {Chat\'e}}, \bibinfo
  {author} {\bibfnamefont {C.}~\bibnamefont {Nardini}},\ and\ \bibinfo {author}
  {\bibfnamefont {A.}~\bibnamefont {Solon}},\ }\bibfield  {title} {\bibinfo
  {title} {Self-organized critical coexistence phase in repulsive active
  particles},\ }\href {https://doi.org/10.1103/PhysRevLett.125.168001}
  {\bibfield  {journal} {\bibinfo  {journal} {Phys. Rev. Lett.}\ }\textbf
  {\bibinfo {volume} {125}},\ \bibinfo {pages} {168001} (\bibinfo {year}
  {2020})}\BibitemShut {NoStop}%
\bibitem [{\citenamefont {Gr\'egoire}\ and\ \citenamefont
  {Chat\'e}(2004)}]{PhysRevLett.92.025702}%
  \BibitemOpen
  \bibfield  {author} {\bibinfo {author} {\bibfnamefont {G.}~\bibnamefont
  {Gr\'egoire}}\ and\ \bibinfo {author} {\bibfnamefont {H.}~\bibnamefont
  {Chat\'e}},\ }\bibfield  {title} {\bibinfo {title} {Onset of collective and
  cohesive motion},\ }\href {https://doi.org/10.1103/PhysRevLett.92.025702}
  {\bibfield  {journal} {\bibinfo  {journal} {Phys. Rev. Lett.}\ }\textbf
  {\bibinfo {volume} {92}},\ \bibinfo {pages} {025702} (\bibinfo {year}
  {2004})}\BibitemShut {NoStop}%
\bibitem [{\citenamefont {Tjhung}\ \emph {et~al.}(2018)\citenamefont {Tjhung},
  \citenamefont {Nardini},\ and\ \citenamefont {Cates}}]{PhysRevX.8.031080}%
  \BibitemOpen
  \bibfield  {author} {\bibinfo {author} {\bibfnamefont {E.}~\bibnamefont
  {Tjhung}}, \bibinfo {author} {\bibfnamefont {C.}~\bibnamefont {Nardini}},\
  and\ \bibinfo {author} {\bibfnamefont {M.~E.}\ \bibnamefont {Cates}},\
  }\bibfield  {title} {\bibinfo {title} {Cluster phases and bubbly phase
  separation in active fluids: Reversal of the ostwald process},\ }\href
  {https://doi.org/10.1103/PhysRevX.8.031080} {\bibfield  {journal} {\bibinfo
  {journal} {Phys. Rev. X}\ }\textbf {\bibinfo {volume} {8}},\ \bibinfo {pages}
  {031080} (\bibinfo {year} {2018})}\BibitemShut {NoStop}%
\bibitem [{\citenamefont {Kurzthaler}\ \emph {et~al.}(2016)\citenamefont
  {Kurzthaler}, \citenamefont {Leitmann},\ and\ \citenamefont
  {Franosch}}]{kurzthaler2016intermediate}%
  \BibitemOpen
  \bibfield  {author} {\bibinfo {author} {\bibfnamefont {C.}~\bibnamefont
  {Kurzthaler}}, \bibinfo {author} {\bibfnamefont {S.}~\bibnamefont
  {Leitmann}},\ and\ \bibinfo {author} {\bibfnamefont {T.}~\bibnamefont
  {Franosch}},\ }\bibfield  {title} {\bibinfo {title} {Intermediate scattering
  function of an anisotropic active brownian particle},\ }\href@noop {}
  {\bibfield  {journal} {\bibinfo  {journal} {Scientific reports}\ }\textbf
  {\bibinfo {volume} {6}},\ \bibinfo {pages} {1} (\bibinfo {year}
  {2016})}\BibitemShut {NoStop}%
\bibitem [{\citenamefont {Kurzthaler}\ and\ \citenamefont
  {Franosch}(2017)}]{kurzthaler2017intermediate}%
  \BibitemOpen
  \bibfield  {author} {\bibinfo {author} {\bibfnamefont {C.}~\bibnamefont
  {Kurzthaler}}\ and\ \bibinfo {author} {\bibfnamefont {T.}~\bibnamefont
  {Franosch}},\ }\bibfield  {title} {\bibinfo {title} {Intermediate scattering
  function of an anisotropic brownian circle swimmer},\ }\href@noop {}
  {\bibfield  {journal} {\bibinfo  {journal} {Soft Matter}\ }\textbf {\bibinfo
  {volume} {13}},\ \bibinfo {pages} {6396} (\bibinfo {year}
  {2017})}\BibitemShut {NoStop}%
\bibitem [{\citenamefont {Takatori}\ and\ \citenamefont
  {Brady}(2015)}]{PhysRevE.91.032117}%
  \BibitemOpen
  \bibfield  {author} {\bibinfo {author} {\bibfnamefont {S.~C.}\ \bibnamefont
  {Takatori}}\ and\ \bibinfo {author} {\bibfnamefont {J.~F.}\ \bibnamefont
  {Brady}},\ }\bibfield  {title} {\bibinfo {title} {Towards a thermodynamics of
  active matter},\ }\href {https://doi.org/10.1103/PhysRevE.91.032117}
  {\bibfield  {journal} {\bibinfo  {journal} {Phys. Rev. E}\ }\textbf {\bibinfo
  {volume} {91}},\ \bibinfo {pages} {032117} (\bibinfo {year}
  {2015})}\BibitemShut {NoStop}%
\end{thebibliography}%

\end{document}